\documentclass[%
 reprint,
groupedaddress,
 amsmath,amssymb,
 aps,
]{revtex4-1}

\usepackage{graphicx}
\usepackage{dcolumn}
\usepackage{bm}

\begin{document}

\preprint{APS/PD10071}

\title{Viewpoint: Pushing Tensor Networks to the Limit}

\author{Anastasiia A.~Pervishko}
 \affiliation{Skoltech's Deep Quantum Laboratory\\Skolkovo Institute of Science and
Technology, 3 Nobel Street, Moscow 143026, Russia}
 
\author{Jacob D.~Biamonte}
\affiliation{Skoltech's Deep Quantum Laboratory\\Skolkovo Institute of Science and
Technology, 3 Nobel Street, Moscow 143026, Russia}
 
\begin{abstract}
This {\it Physics} viewpoint considers recent work by Tilloy and Cirac \cite{TC19}---those authors overcame several past limitations in the generalization of tensor networks to the continuum and proposed a new class of continuous tensor network  states (cMPS) which apply to spatial dimensions of two and higher.  \\

\noindent {\bf An extension of tensor networks---mathematical tools that simplify the study of complex quantum systems—could allow their application to a broad range of quantum field theory problems.}
\end{abstract}  

                              
\maketitle


Not long after the birth of quantum mechanics, Paul Dirac and others postulated that, in principle, quantum mechanics could predict any desired property of matter \cite{Dirac:1930:PQM}. That is, provided one can solve the relevant quantum equations. Often, however, these equations are fiendishly difficult or impossible to solve, as is the case, for instance, for strongly correlated electron systems and other systems in which many-body interactions play an important role. Solving such many-body problems could help us find new high-temperature superconductors, design quantum computing architectures, or describe exotic phase transitions. So-called tensor networks, mathematical tools introduced decades ago, have been widely successful in simplifying the treatment of complex quantum systems. So far, however, these tools could only tackle quantum systems in spatial dimensions higher than 1 by discretizing them—representing them in the form of a discrete lattice. Such a representation can be inadequate for numerous many-body problems. Now, Antoine Tilloy and Ignacio Cirac at the Max Planck Institute for Quantum Optics in Germany have extended tensor networks so that they can represent continuous systems in any spatial dimension (including 2D and 3D) \cite{TC19}. This result may allow researchers to apply tensor networks to a wide class of problems in quantum field theory.

It has long been argued that the equations of quantum mechanics require computational resources far exceeding the capacity of any classical computer. This is due to the “curse of dimensionality”: the number of parameters required to simulate a quantum system using a classical computer appears to scale exponentially with the number of particles that the system contains. As Feynman famously argued, quantum systems themselves could become computational resources: A quantum computer could simulate a quantum system without experiencing the exponential slowdown that a classical computer would. Alas, building a quantum computer that can solve a problem faster than a classical one remains a challenge. But there may be other ways to circumvent the exponential scaling problem by simplifying the quantum problem. A promising approach is offered by tensor networks—elegant mathematical methods that have already solved many important model quantum systems.

Tensor networks provide a simplifying description of a quantum system by representing its wave function through a network of interconnected building blocks called elementary tensors \cite{Orus14, 2011JSP...145..891E, 2013arXiv1308.3318E, 2017arXiv170800006B}. Each tensor is represented by a shape (a square, oval, or triangle) linked by wires. As established by extensive research, these networks provide highly accurate encodings of the relevant quantum properties, including quantum entanglement. Importantly, tensor networks come with a diagrammatic language—much like the renowned Feynman diagrams—which can guide their use through visual intuition.

The computational advantage stems from the fact that the network approximates a complicated quantum state with a simpler structure. In essence, tensor networks can be thought of as a sort of data compression protocol that preserves only those properties of the quantum state that are sufficient to describe its behavior. This compression can dramatically reduce the growth of computational complexity with the scale of the system. Exploiting this advantage, researchers have developed powerful numerical methods based on tensor networks, such as matrix product states—which describe strongly correlated systems in 1D—and projected entangled-pair states \cite{Verstr04} and the multiscale entanglement renormalization ansatz \cite{Vidal07}—which are widely used in 2D.

To apply the above-mentioned methods, the quantum system has to be rewritten in the form of a discrete lattice of elementary tensors. This is an ideal representation for systems made of arrangements of localized spins, like the 1D and 2D spin lattices of the Ising models used to describe magnetic interactions. For other quantum systems, this imposes a possibly unwanted lattice symmetry and can lead to errors. A challenge is thus to generalize tensor networks to the continuum case. Headway was made in 2010 when Frank Verstraete and Cirac extended 1D matrix product states to the continuum limit by letting the lattice spacing tend towards zero while rescaling the tensors \cite{VC10}. The duo showed that the extended formalism can be generally applied to quantum field theories, showing that it can determine the ground states of systems in 1D.

Tilloy and Cirac now further generalize continuum tensor networks to dimensions larger than one. Considering bosonic quantum fields in 2D or 3D, they show that a class of continuous tensor network states can be obtained as the limit of the discrete tensor network representations (Fig.~1). Such states have two equivalent representations, as a functional integral or as an operator. The authors prove that these states can be used as an ``ansatz,” from which one can compute, through a variational approach, expressions for both the n-particle wave function and the correlation functions.

Importantly, they show that the continuous extensions of tensor networks maintain some of the properties that make discrete tensor networks such useful numerical and analytical tools. Specifically, they have the same ``expressiveness”—the ability to approximate any state of the system—and satisfy the same invariance under gauge transformations that don’t alter the state of the system.

Further work will need to address a number of open questions. For example, the authors show that the correlation function of the system can be exactly derived only for continuous tensor network states that can be represented in a Gaussian basis. This is a limited class of states, for which analytical solutions are already known. The approach will have to be generalized to the non-Gaussian case. It will also be interesting to understand if continuous tensor network states obey, similarly to their discrete counterparts, the so-called area law—a fundamental scaling law for the entanglement of a system.

The jury is still out on whether this new continuum extension of tensor network states will lead to new physics. But there is certainly hope that it will provide a range of analytic techniques suitable for tackling continuous systems (such as exact Gaussian functional integrals, saddle-point approximations, or diagrammatic expansions). And the theorists’ approach lays a promising foundation for further studies of quantum field theories using tensor-network-based techniques.\\

The present viewpoint covered the research in \cite{TC19} and was published as \cite{viewpoint}.


\onecolumngrid 

\begin{figure}[h]
\includegraphics[]{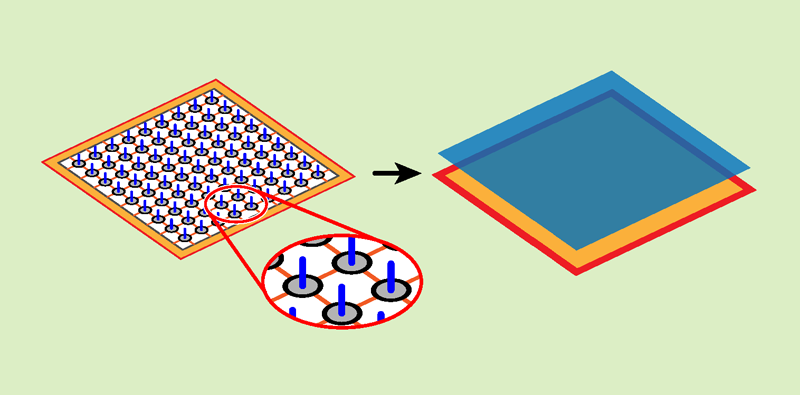} 
\caption{Tilloy and Cirac have extended the application of tensor networks from a 2D lattice case (left) to a continuous case (right) by replacing a sum over discrete indices with a functional integral. (APS/Alan Stonebraker)} 
\end{figure} 

\end{document}